\documentclass[aip,jcp,preprint]{revtex4-1}
\usepackage{graphicx}
\usepackage{amsmath}
\usepackage{amssymb} 
\usepackage{mhchem}
\usepackage[usenames, dvipsnames]{color}

\begin{document}

\title{Like-charge attraction between metal nanoparticles in a 1:1 electrolyte solution}

\author{Alexandre P. dos Santos}
\email{alexandre.pereira@ufrgs.br}
\affiliation{Instituto de F\'isica, Universidade Federal do Rio Grande do Sul, Caixa Postal 15051, CEP 91501-970, Porto Alegre, RS, Brazil.}

\author{Yan Levin}
\email{levin@if.ufrgs.br}
\affiliation{Instituto de F\'isica, Universidade Federal do Rio Grande do Sul, Caixa Postal 15051, CEP 91501-970, Porto Alegre, RS, Brazil.}

\begin{abstract}

We calculate the force between two spherical {\it metal} nanoparticles of charge $Q_1$ and  $Q_2$ in a dilute 1:1 electrolyte solution.  Numerically solving the non-linear Poisson-Boltzmann equation, we find that metal nanoparticles with the same sign of charge can attract one another.  This is
fundamentally different from what is found for like-charged, non-polarizable,  colloidal particles, the two body interaction potential for which is always repulsive inside a dilute 1:1 electrolyte.  Furthermore, existence of like-charge attraction between spherical metal nanoparticles is even more surprising in view of the result that such attraction is impossible between parallel metal slabs, showing the fundamental importance of curvature.
To overcome a slow convergence of the numerical solution of the full non-linear Poisson-Boltzmann equation, we  developed a modified Derjaguin
approximation which allows us to accurate and rapidly calculate the interaction potential between two metal nanoparticles, or between a metal nanoparticle and a phospholipid membrane.     

\end{abstract}

\maketitle

\newpage

% \section{Introduction}

% \section{Model and Theory}

Metal nanoparticles  suspended in an electrolyte solution have attracted a lot of attention for
various applications~\cite{SeMa03,DaAs04,ArBa06,GhKi08,WaKo11,GiDo17,EtRo18,PeRo18,GeRa18,GiMa18}. Because of their strong affinity for biological surfaces and compatibility with immune
system~\cite{CoMw05}, gold nanoparticles are being used for cancer treatment and drug delivery~\cite{RoGi06,DyKh11,GiSe10}. They have also found applications in catalysis~\cite{LoJa04,Th07} and optics~\cite{WaBa06,AhKi17}. Unfortunately our theoretical understanding of the interactions between metal nanoparticles inside an electrolyte solution is rather limited.  Gold nanoparticles are often synthesized  using citrate as a stabilizing agent~\cite{MoCe09,DoBa13}, resulting in a polydisperse suspension of negatively charged nanoparticles at pH 7.
When such particles are in vicinity of one another, in addition to the direct Coulomb force between the two particles, there is an additional interaction arising from the induced charge on the metal 
cores. The induced charge is non-uniformly distributed over the metal cores, but its net amount is zero for each particle.  As the two nanoparticles approach one another, both the surface charge distribution and the electrostatic potential on each particle change with the distance of separation.  Most theoretical works on colloidal suspensions
ignore the effects of polarizability and treats the particle surface
charge distribution
as fixed and uniform~\cite{Le02}. There are, however, some recent works which explore effects of charge regulation~\cite{MaBi18} and patchiness~\cite{SiBe12,BeAn13,BaDo15,AdAn17,MaBi18} on the interaction between planar surfaces, the physics behind such systems, however, is quite different from the polarizability effects that we will be interested to explore in the present Letter. Recent computational methods try to mimic the behavior of metallic materials using parametrized Lennard-Jones particles~\cite{HeLi13,GeRa18}. With the exception of metal planar surfaces~\cite{jack,GiDo17}, the direct implementation of proper electrostatic boundary conditions in simulations using Green function methods is very complicated, requiring the use of computationally very demanding  boundary elements methods in order to account for polarization effects~\cite{TySu10,GaWu15}.  
%or spherical~\cite{jack,Li86,LiLu14,PeRo18} surfaces the boundary conditions imposed by Maxwell equations lead to very complicated potentials which depend on infinity sums in general.
%In order to simplify the solution of Poisson equation charged spheres have been modeled with the condition of constant electrostatic potential~\cite{HoHe66,ChCh83}. This approximation works for larger separations, the correct assumption of idealized metallic nanoparticles is that their charges are held constant in suspensions. The calculations become much more complex in this context. We shed light in this detail motived by the fact that gold nanoparticles have been sintetized with amino acids charged functional groups on their surfaces by ligand exchange reactions~\cite{SeMa03,GhKi08}. The stability of tailored charged metallic nanoparticles is not well described in literature.

It is well known that like-charged colloidal particles can attract one another if suspension contains multivalent counterions~\cite{Pa80,GuJo84,AlDa98,LiBo99,SoDe01,MoNe01,Le02,BuAn03,NaNe04,KaDo09,SaTr11,esdm,DoNe18}.  This attraction results from the electrostatic correlations between the double layers of condensed {\it multivalent} counterions~\cite{Le02}.  On the other hand, it is also believed that 
no such attraction is possible in electrolyte solutions with only 1:1 electrolyte~\cite{Ne99,TrRa99} for which correlation effects are negligible and the mean-field Poisson-Boltzmann (PB) equation is almost exact~\cite{Le02}. Absence of like-charge attraction for non-polarizable colloidal particles  has been confirmed  using explicit Monte-Carlo simulations~\cite{CoDo12}. Furthermore, it can be shown explicitly that like-charged parallel metal slabs inside a dilute 1:1 electrolyte always repel one another. 
Contrary to all of the above, in this Letter we will show that two {\it spherical} like-charged  {\it metal} nanoparticles  
can attract one another in a dilute 1:1 electrolyte solution. The surprising attraction is a consequence of the polarization of the metal cores and is similar to the attraction between charged conducting spheres in vacuum~\cite{Le12,Le16}. The polarization induced like-charge attraction should be very important for the interaction between charged gold particles and phospholipid membranes -- a situation of great practical importance in medical applications~\cite{TaBa14,SiMu16}.

\begin{figure}[t]
\begin{center}
\includegraphics[scale=0.2]{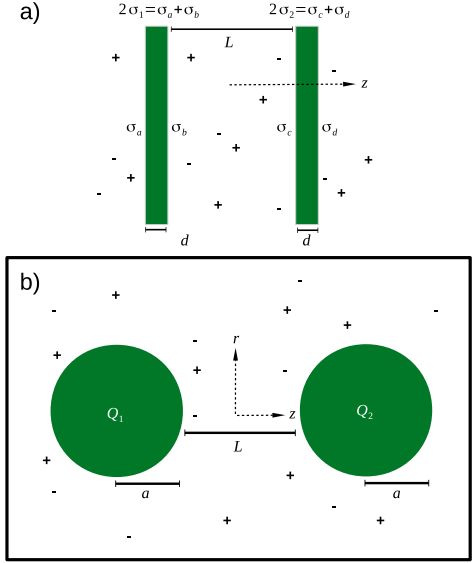}%\vspace{0.2cm}\hspace{0.1cm}
\end{center}
\caption{(a) Two infinite metal slabs of width $d$ and total charge density $2\sigma_1$ and $2\sigma_2$, respectively. $\sigma_{a,b,c,d}$ represent the surface charge densities on the faces of the two slabs.  The values of $\sigma_{a,b,c,d}$ change depending on the separation between the slabs, while the total charge density on each slab remains fixed. (b) Two like-charged spherical metal nanoparticles of charge $Q_1$ and $Q_2$ and radius $a$, separated by a surface-to-surface distance $L$, in an electrolyte solution of concentration $\rho_S$.}
\label{fig1}
\end{figure}

We start by considering the interaction between two parallel infinite metal slabs of width $d$ and total surface charge densities $2\sigma_1$ and $2\sigma_2$, separated by a surface-to-surface distance $L$, as shown in Fig.~\ref{fig1}(a). Both faces of the metal slabs are charged.  The charge on each face will adjust itself so as to minimize the total free energy of the system.  When $L \rightarrow \infty$, both faces of slab 1 will have the same surface charge density $\sigma_1$, and of slab 2, $\sigma_2$.

\begin{figure}[t]
\begin{center}
\includegraphics[scale=0.2]{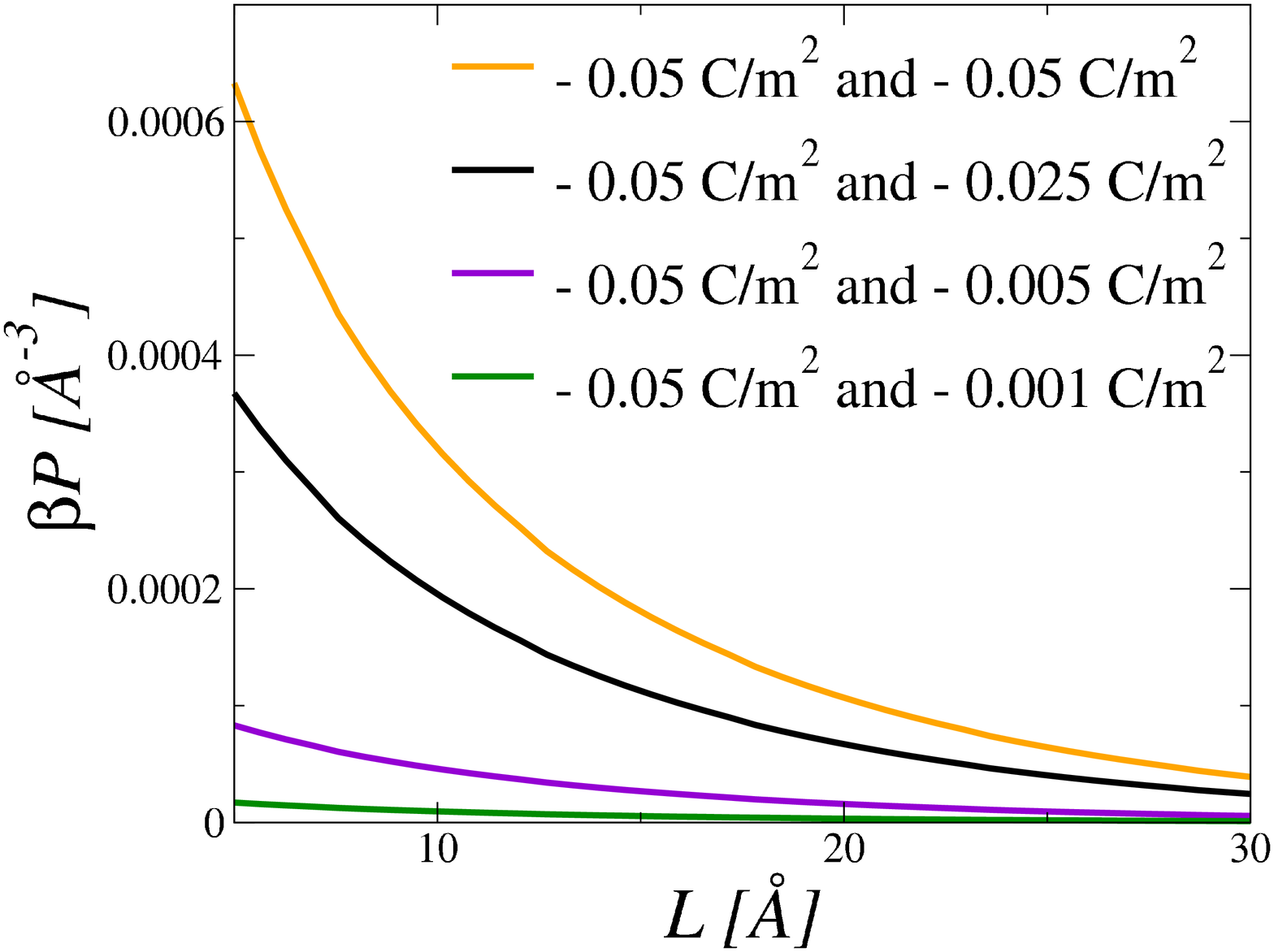}%\vspace{0.2cm}\hspace{0.1cm}
\end{center}
\caption{The pressure between different like-charged metal slabs of width $d=10$\AA, and the charge indicated in the figure, separated by a distance $L$. The pressure is always repulsive, independent of the charge on each slab. The salt concentration is $100~$mM.}
\label{fig2}
\end{figure}

For dilute 1:1 electrolyte solutions, electrostatic correlations between the ions are negligible and the mean field PB equation is quasi exact. To calculate the force between two metal slabs separated by the surface-to-surface distance $L$ 
we must solve the non-linear PB equation
\begin{equation}
\epsilon \nabla^2 \phi = 8 \pi q \rho_S \sinh{[-\beta q \phi]} \ ,
\label{eq1}
\end{equation}
where $\phi$ is the electrostatic potential, $q$ is the elementary charge, $\epsilon$ is the dielectric constant of water, and $\beta=1/k_BT$. The Bjerrum length is defined as $\lambda_B=\beta q^2/\epsilon=7.2~$\AA, the value for water at room temperature.  Inside the metal, the electric field must vanish, so that each slab is an 
equipotential volume.  This means that the contact density of ions on both faces of a slab is identical and the kinetic contribution to the disjoining pressure $P$ must vanish. The pressure  is then determined only by the electric stress  
\begin{eqnarray}
\beta P(L)=\beta \epsilon E_{out}(L)^2/8\pi - \beta \epsilon E_{in}(L)^2/8\pi \ ,
\label{pre}
\end{eqnarray}
where $E_{in}$ and $E_{out}$ are the electric fields at the interior and exterior surfaces of a slab. 
By the superposition, for two  like-charged metal slabs 
$|E_{out}| > |E_{in}|$, so that the pressure will always be repulsive.  This is demonstrated in Fig.~\ref{fig2}, where we have numerically solved the PB equation using using 4$^{th}$ order Runge-Kutta and explicitly calculated the pressure between various like-charged metal slabs.

%%%%%%%%%%%%%%%% figure 2 %%%%%%%%%%%%%%%%%%%%%
%%%%%%%%%%%%% end of figure %%%%%%%%%%%%%%%%%

%%%%%%%%%%%%%%%% figure 1 %%%%%%%%%%%%%%%%%%%%%
%%%%%%%%%%%%% end of figure %%%%%%%%%%%%%%%%%

\begin{figure}[t]
\begin{center}
\includegraphics[scale=0.26]{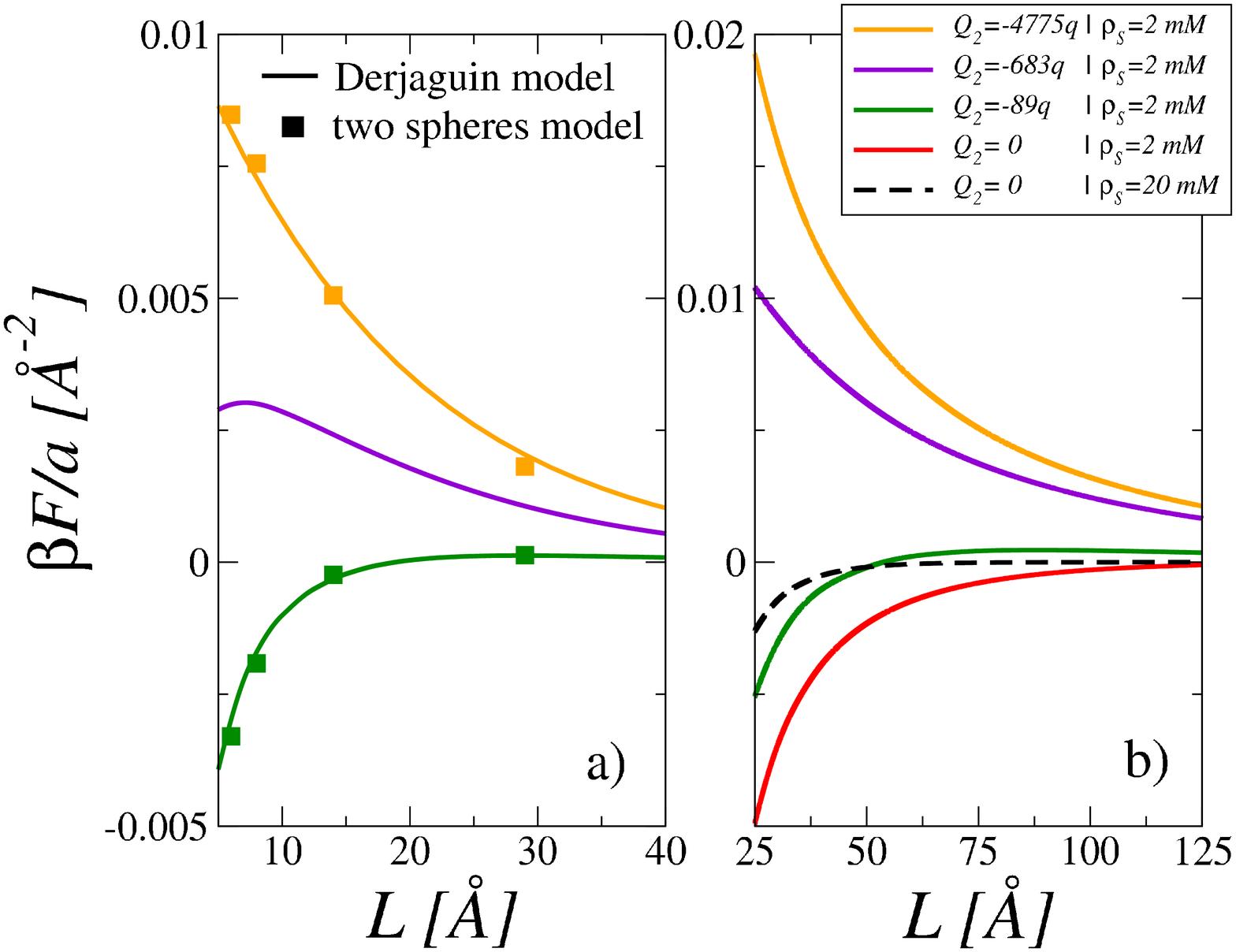}%\vspace{0.2cm}\hspace{0.1cm}
\end{center}
\caption{(a) Electrostatic-entropic force between two like-charged metal nanoparticles of radius $a=50~$\AA\ and charge $Q_1=-67 q$ and $Q_2=-67 q$, $-31 q$ and $-6 q$ --- orange, violet and green curves, respectively --- in electrolyte at $40~$mM. Positive force is repulsive and negative is attractive. 
%The mapping charge densities are $\sigma_1=-27.4~$mC$/$m$^2$ and $\sigma_2=-27.4$, $-12.1$ and $-2.34~$mC$/$m$^2$ for orange, violet and green lines, respectively. 
The squares are forces calculated numerically using PB equation in cylindrical coordinates and Eq.~\ref{eq2}, the lines are calculated using the modified Derjaguin approximation, Eq.~\ref{da}. (b) Electrostatic-entropic force --- calculated using the modified Derjaguin approximation, Eq.~\ref{da} --- between two  like-charged metallic nanoparticles of radius $a=200~$\AA\ and charges $Q_1=-4775q$ and various values of $Q_2$. Salt concentrations as indicated in the figure.}
\label{fig3}
\end{figure}

We next consider two metal nanoparticles depicted in Fig.~\ref{fig1}(b) inside a 1:1 electrolyte solution of concentration $\rho_S$. 
Both particles have radius $a$ and charge $Q_1$ and $Q_2$, respectively.  The surface-to-surface separation is again $L$.   To solve Eq.~\ref{eq1} we now use a relaxation method in cylindrical coordinate system  and define the following boundary conditions: $\phi(\infty,z)=\phi(r,\pm \infty)=\phi'(0,z)=0$, $\phi|_{S_1}=\phi_1$ and $\phi|_{S_2}=\phi_2$ where $\phi_1$ and $\phi_2$ are {\it a priori} unknown electrostatic potentials inside the nanoparticles 1 and 2, respectively. Starting from an initial guess for the values of  $\phi_1$ and $\phi_2$, our algorithm performs a search for the potentials $\phi_1$ and $\phi_2$ until the charge on each nanoparticle --- calculated using the Gauss law, $Q=-\frac{\epsilon}{4\pi} \oint_{S'} {\bf E}\cdot d{\bf S}'$, where  ${\bf E}=-\nabla \phi(r,z)$ is the electric field and $S'$  is the nanoparticle surface  --- agrees with the initially specified value of  $Q_1$ and $Q_2$.  The electrostatic-entropic force per unit volume is ${\bf f}=\nabla \cdot {\bf \Pi}$,
%\begin{eqnarray}
%{\bf f}=\nabla \cdot {\bf \Pi}=\frac{\partial \Pi_{ki}}{\partial x_k}{\bf \hat e}_i
%\end{eqnarray}
where ${\bf \Pi}$ is the entropic-electromagnetic stress tensor 
\begin{eqnarray}
\Pi_{ij}&&=-p(r,z)\delta_{ij}+ \nonumber \\
&&\frac{\epsilon}{4 \pi}\left[E_i(r,z) E_j(r,z) - \frac{1}{2} E^{2}(r,z)\delta_{ij}\right].
\end{eqnarray}
The kinetic pressure is $p(r,z)= k_B T \rho_S (e^{-\beta q\phi(r,z)}+e^{\beta q\phi(r,z)})$, and  $E(r,z)$ and $E_i(r,z)$ are the modulus and the components of the electric field, respectively. The force can be expressed in terms of an integral of the stress tensor over an arbitrary surface enclosing one of the particles, $F=\oint \hat {\bf z} \cdot {\bf \Pi} \cdot \hat {\bf n} dA\ .$
%\begin{eqnarray}
%F=\int \frac{\partial \Pi_{kz}}{\partial x_k}dV \ .
%\end{eqnarray}
%which with the help of the divergence theorem can be written in terms of a surface integral that encloses one of the particles,
% \begin{eqnarray}
% \end{eqnarray}
Choosing the boundary surface to be a cylinder of radius $a$ and length $2a$ we obtain
\begin{eqnarray}
\beta F=2 \pi \int_0^a dr\ r \bigg[ \rho_S e^{-\beta q\phi(r,L/2)}+ \rho_S e^{\beta q\phi(r,L/2)}-\nonumber \\
\rho_S e^{-\beta q\phi(r,L/2+2 a)}-\rho_S e^{\beta q\phi(r,L/2+2 a)} + \nonumber \\
\frac{\beta\epsilon}{8 \pi} \Big[ E_r^2(r,L/2) - E_z^2(r,L/2) + \nonumber \\
E_z^2(r,L/2+2a) - E_r^2(r,L/2+2a)\Big]\bigg] + \nonumber \\
2\pi a \int_{L/2}^{L/2+2a} dz\ \frac{\beta\epsilon}{4\pi} E_r(a,z)E_z(a,z) \ ,
\label{eq2}
\end{eqnarray}
where the positive sign of the force signifies repulsion between the nanoparticles. The results of the numerical integration are shown as  symbols in Fig.~\ref{fig3}(a).  We find that the interaction between two like-charged spherical {\it metal nanoparticles} inside a 1:1 electrolyte solution 
can be either attractive or repulsive, depending on their relative charge and electrolyte concentration!
This is quite surprising in view of our previous result showing that like-charge attraction is impossible between parallel metal slabs.  The curvature of nanoparticles, therefore,  plays a fundamental role for existence of like-charge attraction.  

Unfortunately the relaxation method that we developed to calculate the interaction force between two metal nanoparticles is quite expensive of CPU time.  To obtain accurate results requires a very fine mesh, which makes the convergence very slow, in particular for large particles and low salt concentrations. Furthermore if one of the charged objects is non metal, such as say a phospholipid membrane, significant modifications to the algorithm must be made, since in this case the surface of such object will no longer be equipotential.  In order to overcome these difficulties we have developed a modified Derjaguin approximation, which allows us to efficiently calculate the interaction potential between two metal nanoparticles or between a nanoparticle and a charged planar surface.
%%%%%%%%%%%%%%%% figure 2 %%%%%%%%%%%%%%%%%%%%%
%%%%%%%%%%%%% end of figure %%%%%%%%%%%%%%%%%

Derjaguin approximation allows one to calculate the interaction force between spherical particles, if the corresponding expressions are known for the interaction between planar objects.  
Consider two infinite metal slabs of width $d$ and total surface charge densities $2\sigma_1$ and $2\sigma_2$, separated by a surface-to-surface distance $L$, as depicted in Fig.~\ref{fig1}(a). Both faces of each metal slab are charged, with the surface charge on each face  depending on the separation between the slabs, while the {\it total} surface charge on each slab is fixed.  The values of $\sigma_{1,2}$ are not precisely the surface charge densities on the corresponding spherical nanoparticles.  The nanoparticle surface charge density must be renormalized in order to account for the curvature effects.  This is done by demanding that for large $L \rightarrow \infty$, the electrostatic potential of a metal slab should be the same as for the corresponding nanoparticle. This renormalized surface charge will then produce the same electric field in the vicinity of a slab as exists near a spherical nanoparticle.   The surface potential $\phi_{sp}$ of an isolated spherical particle with a surface charge density $\sigma_{sp}$ 
can be easily calculated by numerically solving the PB equation in spherical coordinates.  Once this is known, the corresponding surface charge density on each face of an isolated slab $\sigma_{sl}$, can be calculated
using the analytical solution of PB equation for a charged plane~\cite{coldis},
\begin{equation}
\sigma_{sl}=\sqrt{\frac{2\rho_S \epsilon}{\pi \beta}}\sinh{(\frac{\beta q\phi_{sp}}{2})} \ .
\label{keqa}
\end{equation}
This provides a mapping between the surface charge densities of spherical nanoparticles and of metal slabs, $\sigma_{1,2}$, used in Derjaguin construction. 
%The model can be understood by taking Fig.~\ref{fig1} and restricting the viewing in region $r\rightarrow 0$. 
%Although we have defined a width of each slab to be $2a$, in practice this width is irrelevant and can be set to $0$ since metal interior is equipotential. 
%The total charge on the corresponding spherical nanoparticles is then $Q_{1,2}=4 a^2 \pi \sigma_{1,2}$.
%electrostatic potentials $\phi_{1,2}$ on each nanoparticle when $L \rightarrow \infty$~\cite{coldis}
%\begin{equation}
%\sigma_{1,2}=\sqrt{\frac{2\rho_S \epsilon}{\pi \beta}}\sinh{(\frac{\beta q\phi_{1,2}^n}{2})} \ .
%\label{keqa}
%\end{equation}
%where $\phi_{1,2}^n$ are the electrostatic potentials of the nanoparticles of charges $Q_{1,2}$ and radii $a$. They can be obtained with a simple PB equation in spherical coordinates using Wigner-Seitz cell model~\cite{AlCh84,Le02}. 

In the spirit of Derjaguin approximation, we now discretize 
the spherical surfaces of each nanoparticle into 
parallel planar slabs. If the disjoining pressure $ P(l)$ between the slabs separated by a surface-to-surface distance $l$ is known, the total force between spherical nanoparticles can be calculated as~\cite{coldis},
\begin{equation}
\beta F = \pi a \int_L^\infty P(l)dl \,.
\label{da}
\end{equation}
The expression for $P(l)$ is the same as in Eq.(\ref{pre}).
%\begin{equation}
%\beta P(l) = \rho_S(e^{-\beta q \phi(\bar z;l)}+e^{\beta q \phi(\bar z;l)})-\frac{\beta \epsilon E^2(\bar z;l)}{8 \pi} - 2 \rho_S \ ,
%\end{equation}
%and is independent of an arbitrary position $\bar z$ between the two slabs used to measure it. 
The validity of Derjaguin approximation is restricted to, $L/a<<1$ and $\kappa a>>1$, where $\kappa=\sqrt{8\pi \lambda_B \rho_S}$ is the inverse Debye length~\cite{ChCh83}. For metal nanoparticles there is an additional complication since the slabs belonging to the same nanoparticle must be equipotential.  However, we do not know {\it a priori} what this potential is, since it depends on the separation between the nanoparticles.  However, we do know that the total charge on each nanoparticle is fixed, independent of separation, which means that the total charge on the slabs that compose a nanoparticle must also be conserved.  This
results in two constraints which determine the electrostatic potentials of metal slabs when the nanoparticles are at surface-to-surface separation $L$,
\begin{eqnarray}
\int_{L}^{L+2a} \Big[\sigma_a(l)+\sigma_b(l)-2 \sigma_1 \Big] dl = 0 \ , \nonumber \\ 
\int_{L}^{L+2a} \Big[\sigma_c(l)+\sigma_d(l)-2 \sigma_2 \Big] dl = 0 \ .
\label{const}
\end{eqnarray}
Note that each slab of our modified Derjaguin approximation has  a different surface charge, while all the slabs corresponding to the same nanoparticle have the same electrostatic potential, which changes with $L$. 
To calculate the disjoining pressure, we first guess the value of the electrostatic potential on each slab, $\phi_1^{guess}$ and $\phi_2^{guess}$.  Since the electric field inside the metal slabs is zero,
the surface charge on the two external faces, see Fig.~\ref{fig1}(a), can be calculated analytically from the exact solution of the PB equation~\cite{coldis},
\begin{equation}
\sigma_{a,d}=\sqrt{\frac{2\rho_S \epsilon}{\pi \beta}}\sinh{\left(\frac{\beta q\phi_{1,2}^{guess}}{2}\right)} \ .
\label{keq}
\end{equation}
To calculate the charge on the interior faces, we numerically integrate the one dimensional PB equation using 4$^{th}$ order Runge-Kutta.  The surface charges $\sigma_{b,c}$ can then be obtained
using the electric field and the Gauss law.  The values of $\phi_1^{guess}$ and $\phi_2^{guess}$
are then adjusted until the constraints given by Eqs.~\ref{const} are satisfied.  In practice, this is done using the Newton-Raphson or some alternative root-finding algorithm.

%%%%%%%%%%%%%%%% figure 2 %%%%%%%%%%%%%%%%%%%%%
% \begin{figure}[b]
% \begin{center}
% \includegraphics[scale=0.23]{fig4.eps}%\vspace{0.2cm}\hspace{0.1cm}
% \end{center}
% \caption{Electrostatic-entropic force --- calculated using the modified Derjaguin approximation, Eq.~\ref{da} --- between two similarly charged metallic nanoparticles of radius $a=200~$\AA\ and charges $Q_1=-4775q$ and various values of $Q_2$. Salt concentrations as indicated in the figure.}
% \label{fig4}
% \end{figure}
%%%%%%%%%%%%% end of figure %%%%%%%%%%%%%%%%%
In Fig.~\ref{fig3}(a) the forces calculated using Eq.~\ref{eq2} and Eq.~\ref{da} are compared. The agreement is very good, showing that the modified Derjaguin approach provides an excellent approximation for calculating the force between metal nanoparticles,  with a significant gain in CPU time.  It is now possible to explore the 
parameter space to see the precise conditions which lead to  like-charge attraction, Fig.~\ref{fig3}(b).  The attraction is a consequence of the non-uniform surface charge induced on the metal cores of the nanoparticles.  However, since the total force contains both electrostatic and entropic contributions, there is no simple criterion that one can use to determine the specific conditions for which like-charge attraction will manifest itself.  In Fig.~\ref{fig3}(b), we use the modified Derjaguin approximation to calculate the force between large nanoparticles of radii $a=200~$\AA, in dilute electrolyte solution  ---  conditions for which a direct integration of the non-linear PB equation is very time consuming. Once again for sufficiently different values of $Q_1$ and $Q_2$, like-charge attraction manifests itself.  Furthermore, we observe that for low salt concentrations, attraction can extend to very large distances.  

%%%%%%%%%%%%%%%% figure 2 %%%%%%%%%%%%%%%%%%%%%
\begin{figure}[t]
\begin{center}
\includegraphics[scale=0.2]{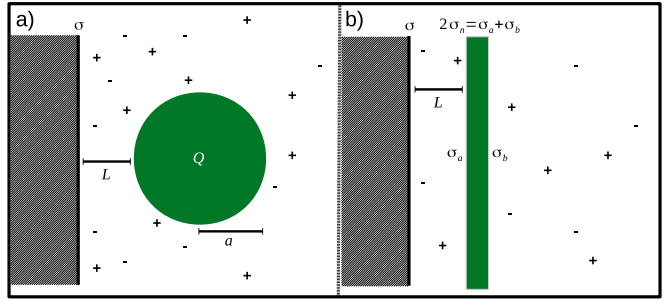}%\vspace{0.2cm}\hspace{0.1cm}
\end{center}
\caption{(a) A metal nanoparticle of charge $Q$ and radius $a$, at surface-to-surface distance $L$ from a charged planar membrane, inside an electrolyte solution. (b) Representation of the modified Derjaguin approximation for this system.}
\label{fig4}
\end{figure}
%%%%%%%%%%%%% end of figure %%%%%%%%%%%%%%%%%

The modified Derjaguin approach introduced in this Letter can also be used to study adsorption of metal nanoparticles with charge  $Q$ to a charged phopholipid membrane with surface charge density $\sigma$, see Fig.~\ref{fig4}. Within the Derjaguin approximation the electric field just outside the membrane is directly determined by the Gauss law, $E(0)= 4 \pi \sigma/\epsilon$, which allows us to easily integrate the 1d PB equation using 4$^{th}$ order Runge-Kutta. The potential 
on the metal slabs is once again determined using the charge conservation condition,
\begin{equation}
\int_{L}^{L+2a} \Big[\sigma_a(l)+\sigma_b-2 \sigma_n \Big] dl = 0 \ ,
\end{equation}
where $2 \sigma_n$ is the renormalized total surface charge on the metal slab, calculated using
Eq.~\ref{keqa}.
%We solve Eq.~\ref{eq1} in 1D Cartesian coordinate system using Runge-Kutta 4$^{th}$ order considering that $E(0)=\frac{4\pi\sigma}{\epsilon}$. The potential at $z=0$ is changed until $\phi(L)=\phi_n$, the electrostatic potential of the metallic surface. For each distance $L$ we calculate $\sigma_a$ for a given $\phi_n$, while $\sigma_b$ is calculated using Eq.~\ref{keq} and $\phi_n$. The following integral must vanish,
%We search for the best $\phi_n$ which satisfy this condition for each distance $L$. The charge density $\sigma_n$ is calculated using Eq.~\ref{keqa} with the electrostatic potential of the isolated nanoparticle of charge $Q$ and radius $a$. It is important to note that $\phi_n$ is free to change while $\sigma_n$ is kept constant.
The electrostatic-entropic force between the membrane and the nanoparticle can be calculated using Eq.~\ref{da}, replacing the prefactor $\pi a$ by $2\pi a$, valid for the interaction of a sphere with a planar surface~\cite{coldis}. The interaction potential can be obtained by integrating the force as a function of separation. To quantitatively study the adsorption of metal nanoparticles to the membrane one must also take into account the dispersion interaction~\cite{Ha37}
\begin{equation}
U_v=-\dfrac{A}{12}\left[ \dfrac{2a}{L}+\dfrac{1}{(1+L/2a)}+2\log{(\dfrac{L/2a}{1+L/2a})} \right] \ ,
\end{equation}
where $A \approx 8.9~k_BT$ is the Hamaker constant characteristic of decane-gold in water at room temperature~\cite{hama}.
%%%%%%%%%%%%%%%% figure 2 %%%%%%%%%%%%%%%%%%%%%
% \begin{figure}[t]
% \begin{center}
% \includegraphics[scale=0.27]{fig5a.eps}%\vspace{0.2cm}\hspace{0.1cm}
% \includegraphics[scale=0.27]{fig5b.eps}%\vspace{0.2cm}\hspace{0.1cm}
% \end{center}
% \caption{Interaction energy between a spherical charged metallic nanoparticle of radius $a=50~$\AA, charges $Q=-67q$ (a), $Q=6q$ (b) and an infinity decane planar surface of charge density $\sigma=-0.26~$C$/$m$^2$ for various electrolyte concentrations.}
% \label{fig5}
% \end{figure}
%%%%%%%%%%%%% end of figure %%%%%%%%%%%%%%%%%
% We observe that in case (a) the increase of electrolyte concentration screens the repulsive electrostatic interaction and can lead the adsorption of the metallic spherical nanoparticle on %the charged dielectric surface by vdW forces. In (b) the electrolyte concentration addition allows the screening of the attractive electrostatic interaction. The superior limit of attraction %is then given by the vdW force.

%%%%%%%%%%%%%%%% figure 2 %%%%%%%%%%%%%%%%%%%%%
\begin{figure}[h]
\begin{center}
\includegraphics[scale=0.21]{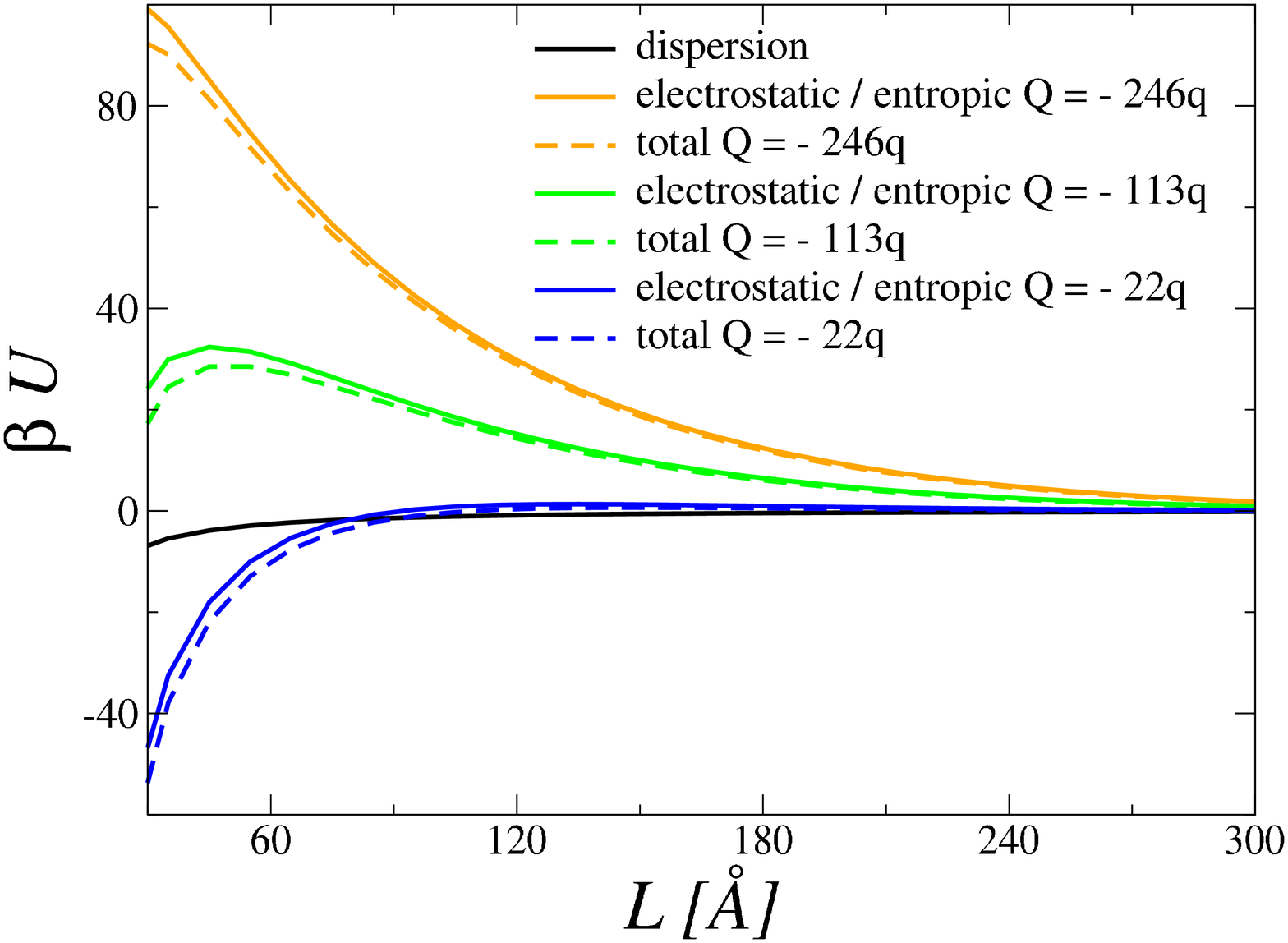}\vspace{-0.2cm}%\hspace{0.1cm}
\end{center}
\caption{Interaction potentials between a spherical metal nanoparticle of radius $a=200~$\AA\ and charge $Q$, indicated in the figure, and a membrane of charge density $\sigma=-0.26~$C$/$m$^2$. The electrolyte concentration is $2~$mM. Solid curves are the electrostatic-entropic potentials for different nanoparticle charges, while the dashed curves are the total interaction potentials, which also include the van der Waals dispersion interaction. The membrane was modeled as a decane, with Hamaker constant $A \approx 8.9~k_BT$.}
\label{fig5}
\end{figure}
%%%%%%%%%%%%% end of figure %%%%%%%%%%%%%%%%%
We now explore the interaction between gold nanoparticles of radius $a=200~$\AA\ and negative charge $Q$, with a like-charged phospholipid membrane of surface charge density $\sigma=-0.26~$C$/$m$^2$, in a dilute electrolyte solution of $2~$mM. We see that strongly charged
nanoparticles are repelled from the surface.  However when the modulus of $Q$ is not too large
the interaction becomes attractive at sufficiently short separations, see Fig.~\ref{fig5}.  As the modulus of the charge decreases, the range of like-charge attraction increases. Fig.~\ref{fig5} also shows that for these low salt concentrations, the total particle-membrane interaction potential is dominated by the electrostatic-entropic contribution, with the dispersion potential being negligible.
If salt concentration is increased, the electrostatic contribution will become screened and the total potential will be dominated by the dispersion interaction. Knowledge of the interaction potential between the metal nanoparticles and a phopholipid membrane allows us to easily calculate the 
adsorption isotherms.  This will be explored in the future work.  

In this Letter we have explored like-charge attraction between spherical metal nanoparticles
inside a monovalent electrolyte solution.  Existence of such attraction
is particularly surprising considering that two like-charged parallel metal slabs always repel one another, demonstrating the importance of curvature for this counterintuitive result.  We have used two methods to explore interaction between nanoparticles  -- a direct numerical integration of the full non-linear PB equation in cylindrical coordinates, and a newly introduced modified Derjaguin approximation.  
Both approaches provide identical results, but the modified  Derjaguin approximation leads to orders of magnitude gain in CPU time. We have also used the modified  Derjaguin approximation to study the adsorption of charged metal nanoparticles to biological membranes. The new theory provides an efficient way to calculate the adsorption isotherms important in various medical applications. It can also be used to study stability of dispersions and to explore heterogeneous coagulation of suspensions of metal nanoparticles.
% \section{Conclusions}

% \section{Acknowledgments}

YL would like to acknowledge very useful conversations with Renato Pakter about numerical methods.  
This work was partially supported by the CNPq, INCT-FCx, and by the US-AFOSR under the grant FA9550-16-1-0280.

\bibliography{ref.bib}

\end{document}